\documentclass[11pt,twoside]{article}
\usepackage{graphicx}
\usepackage{amsmath}
\usepackage{amssymb}
\usepackage{cite}

\begin{document}

\pagestyle{plain}
\newcommand{\pst}{\hspace*{1.5em}}
\setcounter{footnote}{0} \setcounter{equation}{0}
\setcounter{figure}{0} \setcounter{table}{0} \setcounter{section}{0}

\newcommand{\be}{\begin{equation}}
\newcommand{\ee}{\end{equation}}
\newcommand{\bm}{\boldmath}
\newcommand{\ds}{\displaystyle}
\newcommand{\bea}{\begin{eqnarray}}
\newcommand{\eea}{\end{eqnarray}}
\newcommand{\ba}{\begin{array}}
\newcommand{\ea}{\end{array}}
\newcommand{\arcsinh}{\mathop{\rm arcsinh}\nolimits}
\newcommand{\arctanh}{\mathop{\rm arctanh}\nolimits}
\renewcommand{\thefootnote}{\fnsymbol{footnote}}
\newcommand{\bc}{\begin{center}}
\newcommand{\ec}{\end{center}}
\newcommand{\degree}{$^\circ$}
\newcommand{\blackcircle}{\begin{picture}(6,12)\put(3,3){\circle*{6}}\end{picture}}

\setcounter{footnote}{0} \setcounter{equation}{0}
\setcounter{figure}{0} \setcounter{table}{0} \setcounter{section}{0}

\thispagestyle{plain}
\label{sh}

\noindent\textbf{\Large \tabcolsep=0mm
\begin{tabular}{c}
\rule{175mm}{0mm}\\[-6mm]
WIGNER FUNCTIONS AND SPIN TOMOGRAMS
\\[-1mm]
FOR QUBIT STATES
\end{tabular}
}

\bigskip

\begin{center} {\bf
Peter Adam,$^1$ Vladimir A. Andreev,$^{2\,*}$ Iulia Ghiu,$^3$
Aurelian Isar,$^4$ Margarita A. Man'ko,$^2$ and Vladimir I.
Man'ko$^2$ }\end{center}

\medskip

\begin{center}
{\it $^1$Institute for Solid State Physics and Optics\\
Wigner Research Centre for Physics, Hungarian Academy of Sciences\\
H-1525 Budapest, P.O. Box 49, Hungary}

\smallskip

{\it $^2$P.~N.~Lebedev Physical Institute, Russian Academy of Sciences\\
Leninskii Prospect 53, Moscow 119991, Russia}

\smallskip

{\it $^3$Centre for Advanced Quantum Physics\\
Department of Physics, University of Bucharest\\
P.O. Box MG-11, R-077125  Bucharest-M\u{a}gurele, Romania}

\smallskip

{\it $^4$Horia Hulubei National Institute for Research and Development\\
in Physics and Nuclear Engineering\\
P.O. Box MG-6, Bucharest-M\u{a}gurele, Romania}

\smallskip

$^*$Corresponding author e-mail:~~~andrvlad\,@\,yandex.ru\\
e-mails:~~~adam\,@\,optics.szfki.kfki.hu~~~iulia.ghiu\,@\,g.unibuc.ro\\
isar\,@\,theory.nipne.ro~~~mmanko\,@\,sci.lebedev.ru~~~manko\,@\,sci.lebedev.ru

\end{center}

\begin{abstract}\noindent
We establish the relation of the spin tomogram to the Wigner
function on a discrete phase space of qubits. We use the quantizers
and dequantizers of the spin tomographic star-product scheme for
qubits to derive the expression for the kernel connecting Wigner
symbols on the discrete phase space with the tomographic symbols.
\end{abstract}

\medskip


\section{Introduction}
\pst The Wigner function is a powerful tool for representing quantum
states and treating quantum-mechanical problems. It is a
quasiprobability distribution and it has also been generalized for
discrete quantum systems \cite{JSC,WKW,AV,Gibbons,KRS}. Properties
of quasidistributions in a finite Hilbert space have been studied in
the literature \cite{PT,CMM,FM1}. The quasidistributions can be
associated with mutually unbiased bases~(MUB)~\cite{MUB1,MUB2}. Also
there exists the construction of tomographic-probability
distributions (spin tomography)~\cite{DM,OMM,AM,AMMS} and
unitary-matrix tomography~\cite{MMSV} describing the quantum states.
The star product of functions~\cite{ST,FFL} is the usual framework
to consider the Wigner function~\cite{EW} and associative product of
functions in the phase space. The star-product approach was
generalized~\cite{MMM} for considering different schemes. The
approach is based on the existence of the so-called dequantizer
$\hat Q(x)$ and quantizer $\hat D(x)$ acting in a Hilbert space and
depending on a collective coordinate $x$ of a point in a manifold.

There exists the geometrical description of quantum states based on
the discrete phase space. The discrete phase space for a quantum
system characterized by a dimension equal to the power of the prime
$d=p^n$ was constructed in \cite{WKW,Gibbons} with the help of $d^2$
points $(x_1,x_2)$, where $x_1$ runs along the horizontal axis and
$x_2$, along the vertical one. A line is described by a subset of
$d$ points. A given set of $d$ parallel lines defines a
striation~\cite{WKW,Gibbons}. Two striations are called mutually
orthogonal if each line of the first striation has exactly one
intersecting point with each line of the second
striation~\cite{Wootters-Found}. There are $d+1$ mutually unbiased
striations.

The correspondence between a line $\lambda $ and a quantum state is
determined by the function $Q$~\cite{Gibbons}, namely, $Q(\lambda)$
is the projection operator of the pure state. The discrete Wigner
function was introduced in \cite{Gibbons} and is based on a special
family of Hermitian operators ${\cal A}_\alpha$, which depend on a
point in the discrete phase space. If $\alpha $ is a point in the
discrete phase space, the phase-space point operators are defined as
\begin{equation}
\label{J1}
{\cal A}_\alpha=\sum_{\lambda \ni \alpha}Q(\lambda )-I,
\end{equation}
where the sum is taken over all lines $\lambda $ that contain the
point $\alpha $. Here $I$ is the identity operator. These operators
satisfy $\mbox{Tr}\,{\cal A}_\alpha =1$. The discrete Wigner
function of a quantum state $\rho $ is defined as~\cite{Gibbons}
\begin{equation}
W_\alpha=\mbox{Tr}(\rho {\cal A}_\alpha)/d. \label{wigner}
\end{equation}
The set of Hermitian operators ${\cal A}_\alpha $ is not unique; it
depends on the complete set of mutually orthogonal striations
constructed with the help of mutually unbiased bases. It turns out
that the MUBs are determined by the bases associated with each
striation. Starting from this geometrical description, more results
were obtained for different systems: two
qubits~\cite{KRS,Ghiu-2012}, three
qubits~\cite{Bjork-2007,Ghiu-2013}, and $n$
qubits~\cite{Klimov-2009}. An interesting analogy was made between
the mutually orthogonal striations and Latin
squares~\cite{Wootters-Found,Paterek-R,Paterek-2010} and a more
general concept called supersquares~\cite{Ghiu-matem}. A recent
detailed review presents different constructions of
MUBs~\cite{Durt-2010}.

An algorithm for constructing the discrete Wigner function in the
case of composed systems, whose dimension can be factorized into
prime factors, $d=d_1\ldots d_p$, was proposed in \cite{Bjork-2001}.
In this case, the phase-space point operators can be written as
tensor products of the phase-space point operators of each
subsystem. The discrete Wigner function of two qubits was used for
evaluating the entanglement in \cite{Franco}. The entanglement was
analyzed with the help of the partial transposition criterion and
the local uncertainty relations, which were reformulated in terms of
the discrete Wigner function.

The aim of this work is to associate the discrete Wigner function
construction of \cite{Gibbons} with the star-product
quantizer--dequantizer scheme and find an explicit formula
connecting the tomographic probabilities and the quasidistributions
on the discrete phase spaces, using the elaborated framework of the
star-product schemes~\cite{MMM,IMM}. In this paper, we present the
one-qubit case. We study in detail a concrete example of the qubit
state, using explicit forms of the quantizer and dequantizer
determining the tomographic probability distribution given in
\cite{FM3}.

This paper is organized as follows.

In  Sec.~2, we review a general scheme of the star-product
quantization. In Sec.~3, we give the construction of the Wigner
function for one qubit within the framework of the star-product
scheme. We study the relation of the Wigner functions to qubit
tomograms in Sec.~4. It is worth noting that some aspects of the
problem of connection of quasidistributions and
tomographic-probability distributions of spin states were discussed
in \cite{MS}. We present our conclusions and prospectives in Sec.~5.

\section{General Description of the Star-Product Scheme}
\pst Following \cite{MMM}, we present a general scheme of the
star-product construction. Given a Hilbert space ${\cal H}$ and
operators called the dequantizers $\hat Q(x)$ and quantizers $\hat
D(x)$ acting on the Hilbert space, these operators satisfy the
condition that, for an arbitrary operator $\hat A$, one has
\begin{equation}
\int dx'\, \mbox{Tr} \big( \hat Q(x)\hat D(x')\big) \mbox{Tr} \big(
\hat A\hat Q(x')\big) = \mbox{Tr} \big( \hat A\hat Q(x)\big).
\label{operat}
\end{equation}
In some cases, such an equality can be rewritten as
\begin{equation}
\mbox{Tr}\,\big( \hat Q(x)\hat D(x')\big) =\delta (x-x').
\end{equation}
We consider the manifold point coordinate $x$ as $(q,p)$ for the
standard phase space of an oscillator. Also, in other cases, this
coordinate can contain discrete components. So far we are dealing
with the spin-$j$ tomography $x=(m,\vec{n})$, where $m$ is the spin
projection taking values $-j,-j+1,\ldots,j$, and $\vec{n}=(\cos
\varphi \sin \theta , \sin \varphi \sin \theta , \cos \theta )$ is a
vector determining the direction, in which we obtain the spin
projection $m$.

We introduce the function
\begin{equation}
f_A(x)=\mbox{Tr} \big( \hat A\hat Q(x)\big),
\end{equation}
called the symbol of the operator $\hat A$. Relation~(\ref{operat})
provides the possibility to reconstruct the operator $\hat A$ from
its symbol
\begin{equation}
\hat A=\int f_A(x)\hat D(x)\,dx.
\end{equation}
There exists a dual symbol (see also \cite{Franco}) of the operator
$\hat A$ given as
\begin{equation}
f_A^d(x)=\mbox{Tr}\, \big( \hat A\hat D(x)\big).
\end{equation}
The reconstruction relation reads
\begin{equation}
\hat A=\int f_A^d(x)\hat Q(x)\,dx.
\end{equation}
The mean value of the observable $\hat A$ in the state characterized
by the density operator $\hat \rho $ is
\begin{equation}
\langle\, \hat A\, \rangle = \mbox{Tr}\,\big( \hat \rho \hat
A\big)=\int w_\rho(x)f_A^d(x)\,dx, \qquad w_\rho
(x)=\mbox{Tr}\,\big( \hat \rho \hat Q(x)\big).
\end{equation}

We assume that there exists another pair of operators $\hat {\tilde
Q}(y)$ and $\hat {\tilde D}(y)$ with the properties of dequantizers
and quantizers. Then a new symbol of the operator $\hat A$
\begin{equation}
F_A(y)=\mbox{Tr}\,\big( \hat{\tilde Q}(y)\hat A\big)\label{MA1}
\end{equation}
can be related to the symbol $f_A(x)$ by means of the kernel
\begin{equation}
F_A(y)=\int K(y,x)f_A(x)\,dx, \qquad K(y,x)= \mbox{Tr}\,\big( \hat
{\tilde Q}(y)\hat D(y) \big).\label{MA1}
\end{equation}
Analogously
\begin{equation}
f_A(x)=\int {\cal K}(x,y)F_A(y)\,dy, \qquad {\cal K}(x,y)=
\mbox{Tr}\,\big( \hat Q(x)\hat{\tilde D}(y) \big).\label{MA2}
\end{equation}
The associative product of two symbols, called the star product, is
defined as
\begin{equation}
(f_A*f_B)(x)=f_{AB}(x),
\end{equation}
and it is determined by the integral kernel
\begin{equation}
(f_A*f_B)(x)=\int f_A(x')f_B(x'')K(x',x'',x)\,dx'\,dx'', \qquad
K(x',x'',x)=\mbox{Tr}\big( \hat D(x')\hat D(x'')\hat Q(x)\big).
\label{kernel}
\end{equation}

\section{The Oscillator Phase Space}
\pst For the standard Wigner function $W(q,p)$ of an oscillator
state, one has the quantizer
\begin{equation}
\hat Q(x)\equiv \hat Q(q,p)=2\,\big\{ \exp \big[ 2\alpha \hat
a^\dagger -2\,\alpha ^*\hat a\big] \big\}\hat I,
\end{equation}
where $\hat a$ and $\hat a^\dagger $ are the creation and
annihilation operators $\hat a =(\hat q+i\hat p)/{\sqrt 2}$ and
$\hat a^\dagger =(\hat q-i\hat p/{\sqrt 2})$, respectively, $\hat I$
is the parity operator, i.e., $\hat I \psi (x)=\psi (-x)$, and the
complex number $\alpha =(q+ip)/{\sqrt 2}$.

In the Weyl--Wigner star-product scheme, the quantizer reads
\begin{equation}
\hat D(x)\equiv \hat D(q,p)=\hat Q(q,p)/{2\,\pi }. \label{quant-D}
\end{equation}
The scheme is self-dual due to (\ref{quant-D}).

The Gr\"{o}newald star-product kernel is given by
Eq.~(\ref{kernel}); it is
\begin{equation}
K(q_1,p_1,q_2,p_2,q_3,p_3)=(2\pi)^{-2}\exp\big \{
2i[q_1p_2-q_2p_1+q_2p_3-q_3p_2+q_3p_1-q_1p_3]\big\}.
\end{equation}
If the symbols of two observables $\hat A$ and $\hat B$ are given as
functionals $A(q,p)$ and $B(q,p)$ in the oscillator phase space, the
Weyl symbol $f_{AB}(q,p)$ of the product $AB$ is given by the
integral
\begin{equation}
f_{AB}(q,p)=\int (2\pi)^{-2}f_{A}(q_1,p_1)f_{B}(q_2,p_2)\exp\big \{
2i[q_1p_2-q_2p_1+q_2p_3-q_3p_2+q_3p_1-q_1p_3]\big\}.
\end{equation}

\section{Example of the Spin 1/2}
\pst For spin equal to 1/2, the dequantizer $\hat Q(x)\equiv \hat
Q(m,\vec{n})$ has the form~\cite{OMM}
\begin{equation}\label{DO}
\hat Q(m,\vec{n})=U^\dagger|\,m\,\rangle \langle\,m\,|U ,
\end{equation}
where $m=\pm 1/2$, and the unitary matrix reads
\begin{equation}
U=\left( \begin{array}{cc}
\cos\vartheta/2~e^{i({\varphi+\psi})/2}&\sin\vartheta/2~e^{i({\varphi-\psi})/2}\\
-\sin\vartheta/2~e^{i({-\varphi+\psi})/2}&\cos\vartheta/2~
e^{-i({\varphi+\psi})/2}\end{array} \right).\label{UMU}
\end{equation}
The dequantizer can be presented in the form of a 2$\times$2 matrix
as follows~\cite{FM3}:
\begin{equation}
\hat Q(m,\vec{n})=\frac{1}{2}\left( \begin{array}{cc}
1&0\\
0&1
\end{array} \right)
+m \left( \begin{array}{cc}
q_{11}&q_{12}\\
q_{21}&q_{22}
\end{array} \right).
\label{deq-one-qubit}
\end{equation}
The quantizer reads
\begin{equation}
\hat D(m,\vec{n})=\frac{1}{2}\left( \begin{array}{cc}
1&0\\
0&1
\end{array} \right)
+3\,m \left( \begin{array}{cc}
q_{11}&q_{12}\\
q_{21}&q_{22}
\end{array} \right).
\end{equation}
For any qubit state with the density matrix
\begin{equation}
\rho=\left( \begin{array}{cc}
\rho_{11}&\rho_{12}\\
\rho_{21}&\rho_{22}
\end{array} \right)
=\frac{1}{2} \left( \begin{array}{cc}
1+z&x+iy\\
x-iy&1-z
\end{array} \right),
\end{equation}
the tomogram reads
\begin{equation}
w(m,\vec{n})=\mbox{Tr}\,\big[\rho \hat Q(m,\vec{n})\big].
\end{equation}
It is the standard probability distribution of the spin projection
$m$ onto the quantization axes $\vec n$, i.e., it is nonnegative,
$w(m,\vec{n})\geq 0$, and the normalization condition
$~\sum_{m=-1/2}^{1/2}w(m,\vec{n})=1$ holds.

The construction of spin tomograms can be generalized for multiqubit
systems.

For the state of two qubits with the density matrix $\rho(1,2)$, one
has the tomogram
\begin{equation}
w(m_1,m_2,\vec{n}_1,\vec{n}_2)=\mbox{Tr}\,\big[\rho (1,2)\hat
Q(m_1,m_2,\vec{n}_1,\vec{n}_2)\big],
\end{equation}
where
\begin{eqnarray}
\label{TP1} \hat Q(m_1,m_2,\vec{n}_1,\vec{n}_2)=\hat Q_1(m_1,\vec{n}_1)\otimes
\hat Q_2(m_2,\vec{n}_2),
\end{eqnarray}
and $\hat Q_1$ and $\hat Q_2$ are given by (\ref{deq-one-qubit}). It
is the joint probability distribution of two spin projections $m_1$
and $m_2$ onto the quantization axes $\vec n_1$ and $\vec n_2$,
respectively.

Also the quantizer is the tensor product
\begin{eqnarray}
\label{TP2}
\hat D(m_1,m_2,\vec{n}_1,\vec{n}_2)=\hat D_1(m_1,\vec{n}_1)\otimes
\hat D_2(m_2,\vec{n}_2).
\end{eqnarray}

\section{Wigner Function of the One-Qubit State}
\pst To demonstrate our general construction, we consider the
one-qubit state. The density matrix of this state can be presented
in two forms: either as
\begin{eqnarray}
\label{W1} \hat \rho=\left( \begin{array}{cc}
a&ce^{i\xi}\\ce^{-i\xi}&b\end{array} \right),\qquad a+b=1,
\end{eqnarray}
or as
\begin{eqnarray}
\label{W2} \hat \rho=\frac12\left( \begin{array}{cc}
1+z&x+iy\\x-iy&1-z\end{array} \right),\end{eqnarray} corresponding
to the two bases: $A$ and $B$. The explicit forms of basis~$A$ and
basis~$B$ are given below.

We introduce four matrices $\hat{\cal A}_\alpha $ for one qubit,
where the collective index $\alpha(j,k)$ takes the values (0,0),
(1,0), (0,1), and (1,1). The four matrices $\hat{\cal A}_\alpha $
read \begin{eqnarray*} \hat{\cal A}_{0,0}=\left(
\begin{array}{cc}
1&(1-i)/2\\
(1+i)/2&0
\end{array} \right),\qquad
\hat{\cal A}_{0,1}=\left( \begin{array}{cc}
1&(-1+i)/2\\
(-1-i)/2&0
\end{array} \right),\\
\hat{\cal A}_{1,0}=\left( \begin{array}{cc}
0&(1+i)/2\\
(1-i)/2&1
\end{array} \right),\qquad
\hat{\cal A}_{1,1}=\left( \begin{array}{cc}
0&(-1-i)/2\\
(-1+i)/2&1
\end{array} \right).
\end{eqnarray*}
The four matrices $\hat{\cal B}_\alpha $ for one qubit are given by
the transposed matrices $\hat{\cal A}_\alpha $; they are
\begin{eqnarray*} \hat{\cal B}_{0,0}=\hat{\cal A}_{0,0}^T=\left(
\begin{array}{cc}
1&(1+i)/2\\
(1-i)/2&0
\end{array} \right),\qquad
\hat{\cal B}_{0,1}=\hat{\cal A}_{0,1}^T=\left( \begin{array}{cc}
1&(-1-i)/2\\
(-1+i)/2&0
\end{array} \right),\\
\hat{\cal B}_{1,0}=\hat{\cal A}_{1,0}^T=\left( \begin{array}{cc}
0&(1-i)/2\\
(1+i)/2&1
\end{array} \right),\qquad
\hat{\cal B}_{1,1}=\hat{\cal A}_{1,1}^T=\left( \begin{array}{cc}
0&(-1+i)/2\\
(-1-i)/2&1
\end{array} \right).
\end{eqnarray*}

We are looking for the Wigner function of the one-qubit state in
basis~$A$, where its components read
\begin{eqnarray}  
W^A(j,k)=\mbox{Tr}\,\big(\hat \rho \hat{{\cal
A}}_{j,k}\big)/2,\qquad j,k=0,1,\nonumber\\
W^A(0,0)=(1+z+x-y)/4,\qquad W^A(0,1)=(1+z-x+y)/4,\label{W3}\\
W^A(1,0)=(1-z+x+y)/4,\qquad W^A(1,1)=(1-z-x-y)/4,\nonumber
\end{eqnarray}
while in basis~$B$ they are
\begin{eqnarray}   
W^B(j,k)=\mbox{Tr}(\hat \rho \hat{{\cal B}}_{j,k})/2 \qquad j,k=0,1,
\nonumber\\
W^B(0,0)=((1+z+x+y))/4,\qquad W^B(0,1)=(1+z-x-y)/4,\label{BW3}\\
W^B(1,0)=((1-z+x-y))/4,\qquad
W^B(1,1)=(1-z-x+y)/4.\nonumber\end{eqnarray} We have two sorts of
Wigner functions -- the first one is determined by
dequantizer~(\ref{W3}), and the second one is determined by
dequantizer~(\ref{BW3}).

One can easily check that the following reconstruction formulas are
valid:
\begin{equation}
\label{W5} \hat \rho =\sum_{j,k=0}^1 W^A(j,k)\hat{{\cal A}}_{j,k},\qquad \hat
\rho =\sum_{j,k=0}^1 W^B(j,k)\hat{{\cal B}}_{j,k}.
\end{equation}
The components of the Wigner function in basis~$A$ and basis~$B$ are
related as follows:
\begin{equation}
\label{ABW5}
 W^A(i,j)=\frac12\sum_{l,k=0}^1 W^B(l,k)\mbox{Tr}\,\big(\hat{{\cal A}}_{i,j}\hat{{\cal B}}_{l,k}\big).
\end{equation}

The qubit-state tomogram is determined by various
quantizer--dequantizer pairs.

The tomographic dequantizer operator~(\ref{DO}), with unitary
matrix~(\ref{UMU}) and a unit vector  $\vec n$ with components
$(\sin\vartheta\sin\psi, \sin\vartheta\cos\psi, \cos\vartheta)$, has
the explicit matrix form
\begin{eqnarray}
\label{W8} \hat Q\left(1/2,\vartheta,\psi\right)=\frac12\left(
\begin{array}{cc} 1&0\\0&1\end{array} \right)+ \frac12\left( \begin{array}{cc}
\cos\vartheta &\sin\vartheta e^{-i\psi}\\
\sin\vartheta e^{i\psi}&-\cos\vartheta \end{array} \right),\nonumber\\[-2mm]
\\[-2mm]\hat Q\left(-1/2,\vartheta,\psi\right)=\frac12\left(
\begin{array}{cc} 1&0\\0&1\end{array} \right)- \frac12\left( \begin{array}{cc}
\cos\vartheta &\sin\vartheta e^{-i\psi}\\
\sin\vartheta e^{i\psi}&-\cos\vartheta \end{array} \right).\nonumber
\end{eqnarray}
The tomographic quantizer operator $\hat D$ has the matrix form with
matrix elements depending on the coordinates of the unit vector
$\vec n$, namely,
\begin{eqnarray}
\label{W9} \hat D\left(1/2,\vartheta,\psi\right)=\frac12\left(
\begin{array}{cc} 1&0\\0&1\end{array} \right)+ \frac32\left( \begin{array}{cc}
\cos\vartheta &\sin\vartheta e^{-i\psi}\\
\sin\vartheta e^{i\psi}&-\cos\vartheta \end{array} \right),\nonumber\\[-2mm]
\\[-2mm] \hat D\left(-1/2,\vartheta,\psi\right)=\frac12\left(
\begin{array}{cc} 1&0\\0&1\end{array} \right)- \frac32\left( \begin{array}{cc}
\cos\vartheta &\sin\vartheta e^{-i\psi}\\
\sin\vartheta e^{i\psi}&-\cos\vartheta \end{array} \right).
\nonumber\end{eqnarray} In view of formulas~(\ref{MA1}) and
(\ref{MA2}), we obtain the kernels connecting tomograms and Wigner
functions.

After some algebra, we obtain the components of the kernel $~{\rm
Ker}^A(m,\vec n;j,k)={\rm Tr}\,\big(\hat Q(m,\vec n)\hat{{\cal
A}}_{j,k}\big)$ as follows:
\begin{eqnarray}
\label{W11} {\rm
Ker}^A\left(1/2,\vartheta,\psi;0,0\right)=\big(1+\cos\vartheta+\sin\vartheta(\cos\psi+\sin\psi)\big)/2,
\nonumber\\
{\rm
Ker}^A\left(-1/2,\vartheta,\psi;0,0\right)=\big(1-\cos\vartheta-\sin\vartheta(\cos\psi+\sin\psi))/2,\nonumber\\
{\rm
Ker}^A\left(1/2,\vartheta,\psi;0,1\right)=\big(1+\cos\vartheta-\sin\vartheta(\cos\psi+\sin\psi)\big)/2
\nonumber\\
{\rm Ker}
^A\left(-1/2,\vartheta,\psi;0,1\right)=\big(1-\cos\vartheta+\sin\vartheta(\cos\psi+\sin\psi)\big)/2,\nonumber\\[-2mm]
\\[-2mm]
{\rm Ker}^A\left(1/2,\vartheta,\psi;1,0\right)=\big(1-\cos\vartheta+\sin\vartheta(\cos\psi-\sin\psi)\big)/2,\nonumber\\
{\rm
Ker}^A\left(-1/2,\vartheta,\psi;1,0\right)=\big(1+\cos\vartheta-\sin\vartheta(\cos\psi-\sin\psi)\big),\nonumber\\
{\rm
Ker}^A\left(1/2,\vartheta,\psi;1,1\right)=\big(1-\cos\vartheta-\sin\vartheta(\cos\psi-\sin\psi)\big)/2,\nonumber\\
{\rm
Ker}^A\left(-1/2,\vartheta,\psi;1,1\right)=\big(1+\cos\vartheta+\sin\vartheta(\cos\psi-\sin\psi)\big)/2.\nonumber
\end{eqnarray}
In addition, we find the components of the kernel $~\widetilde {\rm
Ker}^A(m,\vec n;j,k)={\rm Tr}\,\big(\hat D(m,\vec n)\hat{{\cal
A}}_{j,k}\big)/2$ connecting the Wigner functions with the dual
tomograms; they read
\begin{eqnarray}
\label{W13} \widetilde {\rm
Ker}^A\left(1/2,\vartheta,\psi;0,0\right)=\big(1+3\cos\vartheta+3\sin\vartheta(\cos\psi+\sin\psi)\big)/4,\nonumber\\
\widetilde {\rm
Ker}^A\left(-1/2,\vartheta,\psi;0,0\right)=\big(1-3\cos\vartheta-3\sin\vartheta(\cos\psi+\sin\psi)\big)/4,\nonumber\\
\widetilde {\rm Ker}^A\left(1/2,\vartheta,\psi;0,1\right)=\big(1+3\cos\vartheta-3\sin\vartheta(\cos\psi+\sin\psi)\big)/4,\nonumber\\
\widetilde {\rm
Ker}^A\left(-1/2,\vartheta,\psi;0,1\right)=\big(1-3\cos\vartheta+3\sin\vartheta(\cos\psi+\sin\psi)\big)/4,\nonumber\\[-2mm]
\\[-2mm]
\widetilde {\rm Ker}^A\left(1/2,\vartheta,\psi;1,0\right)=\big(1-3\cos\vartheta+3\sin\vartheta(\cos\psi-\sin\psi)\big)/4,\nonumber\\
\widetilde {\rm Ker}^A\left(-1/2,\vartheta,\psi;1,0\right)=\big(1+3\cos\vartheta-3\sin\vartheta(\cos\psi-\sin\psi)\big)/4,\nonumber\\
\widetilde {\rm Ker}^A\left(1/2,\vartheta,\psi;1,1\right)=\big(1-3\cos\vartheta-3\sin\vartheta(\cos\psi-\sin\psi)\big)/4,\nonumber\\
\widetilde {\rm
Ker}^A\left(-1/2,\vartheta,\psi;1,1\right)=\big(1+3\cos\vartheta+3\sin\vartheta(\cos\psi-\sin\psi)\big)/4.\nonumber
\end{eqnarray}
Now we show the components of the kerners $~{\rm Ker}^B\big(m,\vec
n;j,k\big)={\rm Tr}\,\big(\hat Q(m,\vec n)\hat{{\cal B}}_{j,k}\big)$
and $~\widetilde{\rm Ker}^B\big(m,\vec n;j,k\big)= {\rm
Tr}\,\big(\hat D(m,\vec n)\hat{{\cal B}}_{j,k}\big)/2$ constructed
with the help of operators $\hat{{\cal B}}_{j,k}$; they are
\begin{eqnarray}
\label{BW11} {\rm
Ker}^B\left(1/2,\vartheta,\psi;0,0\right)=\big(1+\cos\vartheta+\sin\vartheta(\cos\psi-\sin\psi))/2,
\nonumber\\
{\rm
Ker}^B\left(-1/2,\vartheta,\psi;0,0\right)=\big(1-\cos\vartheta-\sin\vartheta(\cos\psi-\sin\psi))/2,
\nonumber\\
{\rm
Ker}^B\left(1/2,\vartheta,\psi;0,1\right)=\big(1+\cos\vartheta-\sin\vartheta(\cos\psi-\sin\psi)\big)/2,
\nonumber\\
{\rm
Ker}^B\left(-1/2,\vartheta,\psi;0,1\right)=\big(1-\cos\vartheta+\sin\vartheta(\cos\psi-\sin\psi)\big)/2,
\nonumber\\[-2mm]
\\[-2mm]
{\rm
Ker}^B\left(1/2,\vartheta,\psi;1,0\right)=\big(1-\cos\vartheta+\sin\vartheta(\cos\psi+\sin\psi)\big)/2,
\nonumber\\
{\rm
Ker}^B\left(-1/2,\vartheta,\psi;1,0\right)=\big(1+\cos\vartheta-\sin\vartheta(\cos\psi+\sin\psi)\big)/2,
\nonumber\\
{\rm
Ker}^B\left(1/2,\vartheta,\psi;1,1\right)=\big(1-\cos\vartheta-\sin\vartheta(\cos\psi+\sin\psi)\big)/2,
\nonumber\\ {\rm
Ker}^B\left(-1/2,\vartheta,\psi;1,1\right)=\big(1+\cos\vartheta+\sin\vartheta(\cos\psi+\sin\psi)\big)/2,\nonumber
\end{eqnarray}
and
\begin{eqnarray}
\label{BW13}
\widetilde {\rm Ker}^B\left(1/2,\vartheta,\psi;0,0\right)=\big(1+3\cos\vartheta+3\sin\vartheta(\cos\psi-\sin\psi)\big)/4,\nonumber\\
\widetilde {\rm
Ker}^B\left(-1/2,\vartheta,\psi;0,0\right)=\big(1-3\cos\vartheta-3\sin\vartheta(\cos\psi-\sin\psi)\big)/4,
\nonumber\\
\widetilde {\rm
Ker}^B\left(1/2,\vartheta,\psi;0,1\right)=\big(1+3\cos\vartheta-3\sin\vartheta(\cos\psi-\sin\psi)\big)/4,
\nonumber\\
\widetilde {\rm
Ker}^B\left(-1/2,\vartheta,\psi;0,1\right)=\big(1-3\cos\vartheta+3\sin\vartheta(\cos\psi-\sin\psi)\big)/4,
\nonumber\\[-2mm]
\\[-2mm]
\widetilde {\rm
Ker}^B\left(1/2,\vartheta,\psi;1,0\right)=\big(1-3\cos\vartheta+3\sin\vartheta(\cos\psi+\sin\psi)\big)/4,
\nonumber\\
\widetilde {\rm
Ker}^B\left(-1/2,\vartheta,\psi;1,0\right)=\big(1+3\cos\vartheta-3\sin\vartheta(\cos\psi+\sin\psi)\big)/4,
\nonumber\\
\widetilde {\rm
Ker}^B\left(1/2,\vartheta,\psi;1,1\right)=\big(1-3\cos\vartheta-3\sin\vartheta(\cos\psi+\sin\psi)\big)/4,
\nonumber\\ \widetilde {\rm
Ker}^B\left(-1/2,\vartheta,\psi;1,1\right)=\big(1+3\cos\vartheta+3\sin\vartheta(\cos\psi+\sin\psi))/4.\nonumber
\end{eqnarray}

\section{Tomograms of the One-Qubit State}
\pst The tomograms of the one-qubit state~(\ref{W1}) are
\begin{eqnarray}
\label{W14} w_1=
\frac12+\frac{(a-b)}2\cos\vartheta+c\sin\vartheta\cos(\psi+\xi)=
\frac12(1+z\cos\vartheta+x\sin\vartheta\cos\psi-y\sin\vartheta\sin\psi),\nonumber\\[-2mm]
\\[-2mm]
w_2=
\frac12-\frac{(a-b)}2\cos\vartheta-c\sin\vartheta\cos(\psi+\xi)=
\frac12(1-z\cos\vartheta-x\sin\vartheta\cos\psi+y\sin\vartheta\sin\psi).\nonumber
\end{eqnarray}
We can reconstruct these tomograms, in view of the kernel $~{\rm
Ker}^A(m,\vec n;j,k)={\rm Tr}\,\big(\hat Q(m,\vec n)\hat{{\cal
A}}_{j,k}\big)$, through the Wigner functions (\ref{W3}) and
(\ref{BW3}). After some algebra, we obtain the following
relationships:
\begin{eqnarray}
\label{W15} w_1= w(1/2,\vartheta,\psi)=\sum_{j,k=0}^1{\rm
Ker}^A(1/2,\vartheta,\psi;j,k)W^A(j,k)= \sum_{j,k=0}^1{\rm
Ker}^B(1/2,\vartheta,\psi;j,k)W^B(j,k),\nonumber\\[-2mm]
\\[-2mm] w_2= w(-1/2,\vartheta,\psi)=\sum_{j,k=0}^1{\rm
Ker}^A(-1/2,\vartheta,\psi;j,k)W^A(j,k)= \sum_{j,k=0}^1{\rm
Ker}^B(-1/2,\vartheta,\psi;j,k)W^B(j,k).\nonumber
\end{eqnarray}
We can also reconstruct tomograms~(\ref{W14}), in view of the
kernels $~{\rm Ker}^B\big(m,\vec n;j,k\big)={\rm Tr}\,\big(\hat
Q(m,\vec n)\hat{{\cal B}}_{j,k}\big)$ and $~\widetilde{\rm
Ker}^B\big(m,\vec n;j,k\big)= {\rm Tr}\,\big(\hat D(m,\vec
n)\hat{{\cal B}}_{j,k}\big)/2$,
through the Wigner functions~(\ref{W3}) and (\ref{BW3}). 
After some algebra, we obtain the following relationships:
\begin{eqnarray}
\label{W16} W^A(j,k)=\frac1{4\pi}\sum_{m=-1/2}^{1/2}\int_0^\pi\int_0^{2\pi}
w\left(m,\vartheta,\psi\right)\widetilde {\rm
Ker}^A(m,\vartheta,\psi;j,k)
\sin\vartheta\, d\vartheta\, d\psi,\nonumber\\[-2mm]
\\[-2mm]
W^B(j,k)=\frac1{4\pi}\sum_{m=-1/2}^{1/2}\int_0^\pi\int_0^{2\pi}
w\left(m,\vartheta,\psi\right)\widetilde {\rm
Ker}^B(m,\vartheta,\psi;j,k)\sin\vartheta \,d\vartheta
\,d\psi.\nonumber
\end{eqnarray}

\section{Conclusions}
\pst In conclusion, we list the main results of our study.

We constructed the Wigner functions of the one-qubit state using the
framework of the star-product quantization scheme and explicit forms
of the quantizer and dequantizer operators. We also constructed the
probability distributions for the one-qubit state applying the
star-product scheme and the pair of quantizer and dequantizer
operators determining the state tomogram. We elaborated the
procedure of finding the relation of the tomograms of the qubit
state to the explicit form of kernels providing the map of the qubit
state tomograms onto the Wigner functions and vice versa.

In our approach, the known formulas determining the Wigner function
of quibit state in terms of operators $\hat A_\alpha$~\cite{Gibbons}
are reformulated as formulas used in the star-product quantization
schemes, where the quantizer--dequantizer operator pair provides an
invertible map of operators onto their symbols. Also the qubit-state
tomograms were mapped onto the Wigner functions, in view of the
procedure based on the tomographic quantizer--dequantizer pair. We
calculated the interwinning kernels connecting the different sorts
of Wigner functions and the qubit tomograms given in terms of
quantizer--dequantizer pairs. The application of the elaborated
scheme to the two-qubit and multiqubit states, using an analogous
approach, will be considered in a future publication.

\section*{Acknowledgments}
\pst This work was performed within the framework of the
collaboration between the Hungarian Academy of Sciences and the
Russian Academy of Sciences on the Problem ``Quantum Correlations,
Decoherence in the Electromagnetic Field Interaction with Matter,
and Tomographic Approach to Signal Analysis'' and between the
Romanian Academy of Sciences and the Russian Academy of Sciences on
the Problem ``Fundamental Aspects of Quantum Optics and Quantum
Correlations in Information Theory.'' I.G. was supported by the
Romanian National Authority for Scientific Research under
Grant~PN-II-ID-PCE-2011-3-1012 for the University of Bucharest. A.I.
acknowledges financial support from the Romanian Ministry of
Education and Research under Project~CNCS-UEFISCDI
PN-II-ID-PCE-2011-3-0083. M.A.M and V.I.M. were partially supported
by the Russian Foundation for Basic Research under Project
No.~11-02-00456\_a, and V.A.V. was partially supported by the
Russian Foundation for Basic Research under Project
No.~11-02-01269\_a.

\end{document}